\documentclass[longbibliography,aps,pre,twocolumn,floatfix,showpacs,a4paper,nofootinbib,amsmath,amssymb,superscriptaddress]{revtex4-1}

\usepackage{graphicx}
\usepackage{color}
\usepackage{dcolumn}
\usepackage{bm}
\usepackage{multirow}
\usepackage{booktabs}
\usepackage{slashbox}
\usepackage{algorithm}
\usepackage{algpseudocode}
\usepackage{fancyvrb}
\usepackage{listings}

\newcommand{\be}{\begin{equation}}
\newcommand{\ee}{\end{equation}}
\newcommand{\ba}{\begin{eqnarray}}
\newcommand{\ea}{\end{eqnarray}}
\newcommand{\bd}{\begin{displaymath}}
\newcommand{\ed}{\end{displaymath}}
\renewcommand{\vec}[1]{\mbox{\boldmath$#1$}}

\newcommand{\cz}{\color{black}}
\newcommand{\czz}{\color{black}}

\newsavebox\CBox
\newcommand\hcancel[2][0.5pt]{%
  \ifmmode\sbox\CBox{$#2$}\else\sbox\CBox{#2}\fi%
  \makebox[0pt][l]{\usebox\CBox}%
  \rule[0.5\ht\CBox-#1/2]{\wd\CBox}{#1}}

\usepackage{ulem} 

\newcommand{\switch}[1]{%
  \begingroup
    \def\tmp{#1}%
    \ifx\tmp\1
      first%
    \else\ifx\tmp\2
      second%
    \else\ifx\tmp\3
      third%
    \else
      unknown%
    \fi\fi\fi
  \endgroup}

\renewcommand{\vec}[1]{\mathbf{\textbf{#1}}}

\begin{document}

\title{Stochastic graph Voronoi tessellation reveals community structure}
\author{Zsolt I. L\'az\'ar} \email{zsolt.lazar@phys.ubbcluj.ro}
\affiliation{Faculty of Physics, Babe\c{s}-Bolyai University, Cluj-Napoca, Romania }
\author{Istv\'an Papp}  
\affiliation{Faculty of Physics, Babe\c{s}-Bolyai University, Cluj-Napoca, Romania }
\author{Levente Varga} 
\affiliation{Faculty of Physics, Babe\c{s}-Bolyai University, Cluj-Napoca, Romania }
\author{Ferenc J\'arai-Szab\'o} 
\affiliation{Faculty of Physics, Babe\c{s}-Bolyai University, Cluj-Napoca, Romania }
\author{D\'avid Deritei}
\affiliation{Faculty of Physics, Babe\c{s}-Bolyai University, Cluj-Napoca, Romania }
\affiliation{Department of Network Science, Central European University, Hungary}
\author{M\'aria Ercsey-Ravasz}\email{ercsey.ravasz@phys.ubbcluj.ro}
\affiliation{Faculty of Physics, Babe\c{s}-Bolyai University, Cluj-Napoca, Romania }
\affiliation{Romanian Institute of Science and Technology, Cluj-Napoca, Romania}

\begin{abstract}
Given a network, the statistical ensemble of its graph-Voronoi diagrams with randomly chosen cell centers exhibits properties convertible into information on the network's large scale structures. 
We define a node-pair level measure called {\it Voronoi cohesion} which describes the probability  for sharing the same Voronoi cell, when randomly choosing $g$ centers in the network. 
This measure provides  information based on the global context (the network in its entirety) a type of information that is not carried by other similarity measures.
We explore the mathematical background of this phenomenon and several of its potential applications.
A special focus is laid on the possibilities and limitations pertaining to the exploitation of the phenomenon for community detection purposes. 

\end{abstract}

\date{\today}


\maketitle

\section{Introduction}\label{S:intro}

Community detection in networks consists in identifying subnetworks with pronounced internal connectivity, {\cz also called modules or clusters. A precise and commonly accepted mathematical definition of communities is still missing.} 
Depending on the size and type of the network, desired type of clusterization, requirements applying to resource consumption, accuracy, reproducibility, etc. a wealth of methods are available in the literature {\cz }\cite{Fortunato10,ClusteringPhysRep2013,newman04,pattern,datamining}. 
{\cz When evaluating the optimality of the clusterization  (community structure) provided by an algorithm one can either use a test function such as Newman's modularity measure \cite{newman04} or alternatively confront the result with that of efficient third party community detection methods.
}

Recently, the possibility of identifying {\cz communities} as graph-Voronoi cells has been explored \cite{Deritei14}.
In analogy with Voronoi diagrams in metric spaces \cite{voronoi} [Fig. \ref{F:voronoi}(a)] by picking a number of $g$ nodes from the network, hereinafter referred to as  {\cz \textit{generator}} nodes,  {\cz \textit{centers}} or  {\cz \textit{seeds}}, the network can be separated into $g$ disjoint subnetworks, henceforth referred to as {\cz \textit{cells}}, each ``centered'' on one of the generator nodes [Fig. \ref{F:voronoi}(b)]. 
The nodes in a cell are closest to the generator associated with the respective cell. 
Inter-node distance is defined as the length of the corresponding shortest path  {\cz measured in terms of number of edges} \cite{graphvoronoi}.  
The major challenge for the method  described in \cite{Deritei14} was the identification of generator nodes {\cz and the choice of an appropriate graph metric} so that the obtained cells gave an optimal representation of the community structure of the network. {\cz A possible translation of the problem into a sociological context is considering  a network of voters with established communities wherein candidates (generator nodes) should be proposed such that their voter pools (graph Voronoi cells) optimally overlap with the communities.}
{\czz In this approach focus is on distances to selected nodes and the algorithm is deterministic, in contrast with other methods focusing on pairwise distance defined via dynamic Markov processes taking place on the network \cite{Pons06, Delvenne10}}.

Here we present a stochastic variant of the {\czz above graph-Voronoi based} method. An ensemble of diagrams are generated wherein each {\cz diagram}  is obtained from a set of generator nodes drawn randomly from the network [Fig. \ref{F:intro}].
{\cz Unlike in \cite{Deritei14} this randomized tessellation does not directly relate cells to communities.
However,} the \textit{co-location} probability of a pair of nodes, i.e., the likelihood of the two nodes to share the same Voronoi cell, henceforth referred to as \textit{Voronoi cohesion}, is larger for  an intra-community pair (two nodes belonging to the same community) than that for inter-community pairs.
From the Voronoi cohesion matrix resulting from a relatively small ensemble {\cz of randomized tessellation} one can efficiently extract the {\cz community} structure of the network.  
As we will show, the number of generator nodes  bears relevance, yet it can be fixed ahead, making the method practically parameter-free. 
This approach appears to be applicable on networks with hierarchical and overlapping community structures as well [Fig. \ref{F:multilevel}]. It could also form the basis of  hierarchical clusterization methods that use node-similarity measures \cite{similarity,hierarchical1,hierarchical2,hierarchical3,hierarchical4,hierarchical5}. 
The advantage of Voronoi cohesion compared to simple local similarity measures is that it carries more complex information based on the global structure of the network.

The paper is organized as follows. 
In Sec. \ref{S:overview}, we describe the method {\cz and apply it to the } example of a benchmark network. 
The mathematical background of these empirical findings is explored in Sec. \ref{S:analytical}. 
In Sec. \ref{S:complexity} we study numerically the relation between the complexity and accuracy of the method. 
In Sec. \ref{S:boost} we present a technique for improving community detection by boosting the contrast in the cohesion matrix.
Simulation results and related statistics obtained from real data sets are covered in Sec. \ref{S:real}. 
The advantages and limitations of the method are discussed in the concluding section.
{\cz  A glossary explaining notions of  graph theory and the terminology used in the present paper is included in the Appendix.}
\section{Stochastic graph Voronoi tessellation}\label{S:overview}
\begin{figure}[htbp]  
\begin{center}
\resizebox{1.0\columnwidth}{!}
{\includegraphics{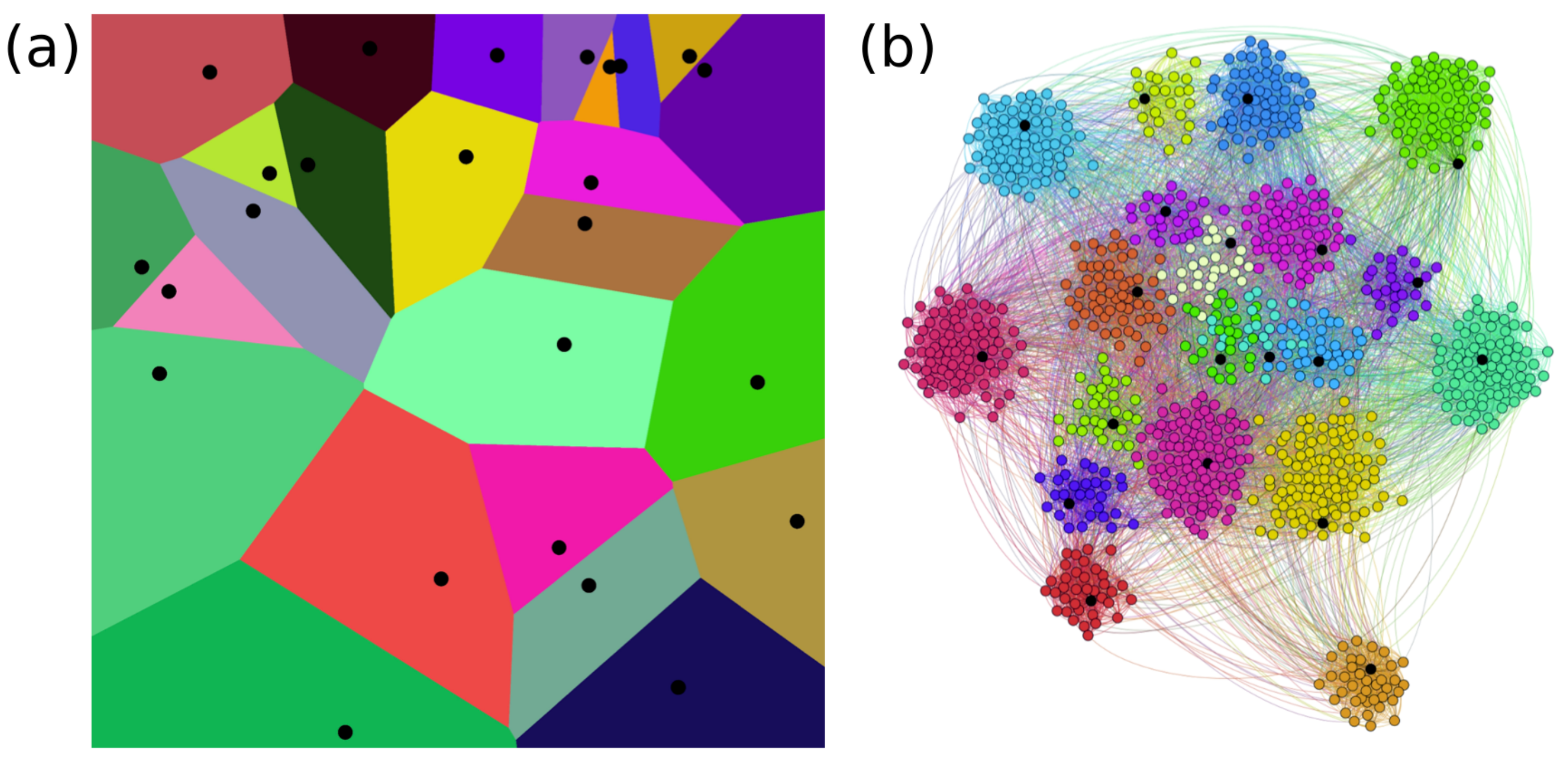}}
\caption{ (color online)  Voronoi tessellation in Euclidean space and on graphs.
(a)Voronoi diagram in 2D Euclidean space
(b) Illustration  of network clustering with Voronoi diagrams. {\cz Layout created by the graph drawing application Gephi using the ForceAtlas2 layout algorithm \cite{gephi}}. Seeds (generators) appear as black dots. {\cz Colors indicate different Voronoi cells, in this case also corresponding well to communities.} 
}

\label{F:voronoi}
\end{center}
\end{figure}
\begin{figure*}[htbp]  
\begin{center}
\resizebox{0.9\textwidth}{!}
{\includegraphics{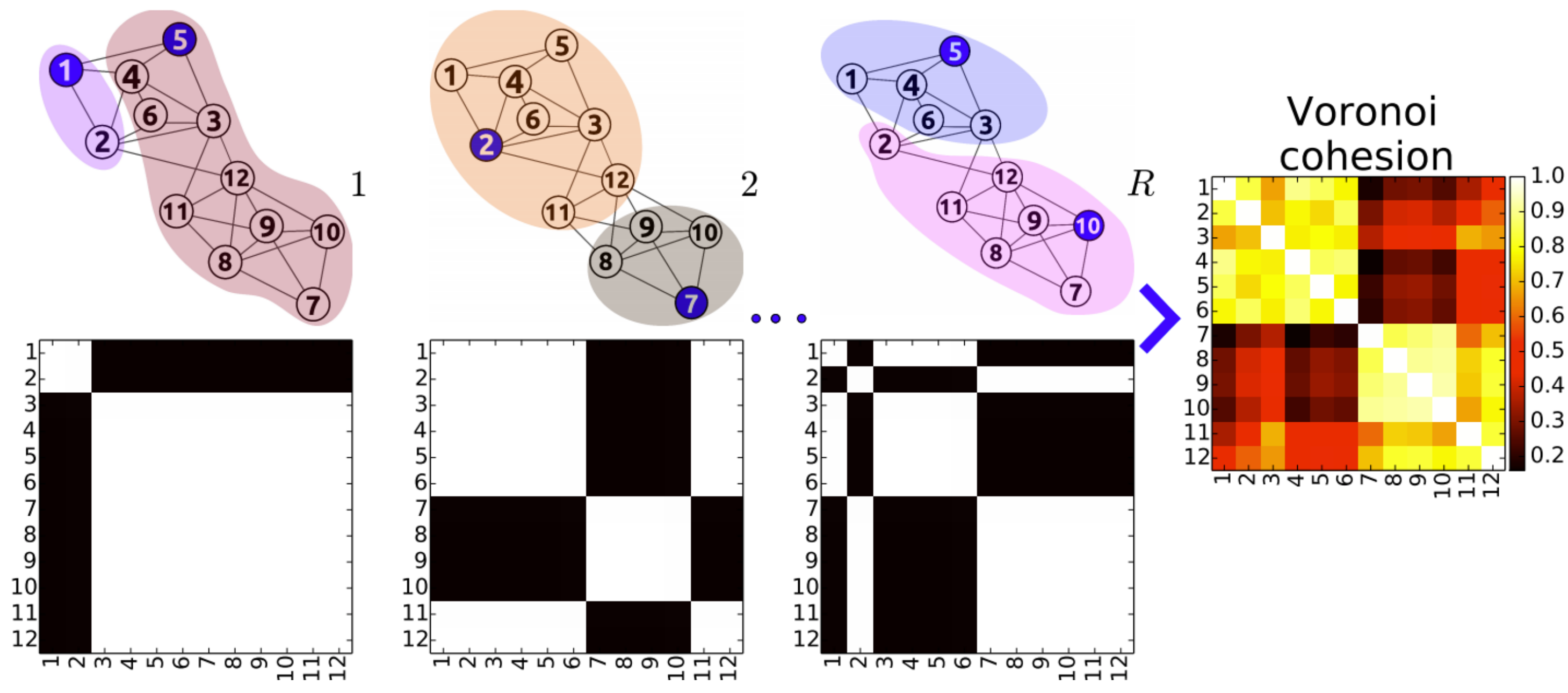}}\,	
\caption{ (color online) First glance at stochastic graph Voronoi tessellation. 
Subfigures show how the Voronoi cohesion map for a twelve node graph would be calculated from an ensemble of $R$ tessellations. The two generator nodes, e.g. (1, 5), (2, 7), ..., (5, 10), are picked randomly producing an ensemble of binary co-location matrices (one ({\cz white}) for same cell, zero ({\cz black}) for different cell node pairs). The ensemble average of the co-location matrices yields the  cohesion map, i.e., probabilities for any node pair to share the same Voronoi cell (lighter {\cz squares} represent higher probabilities, see colorbar).} 

\label{F:intro}
\end{center}
\end{figure*}

\begin{figure}[htbp]  
\begin{center}
\resizebox{0.80\columnwidth}{!}
{\includegraphics{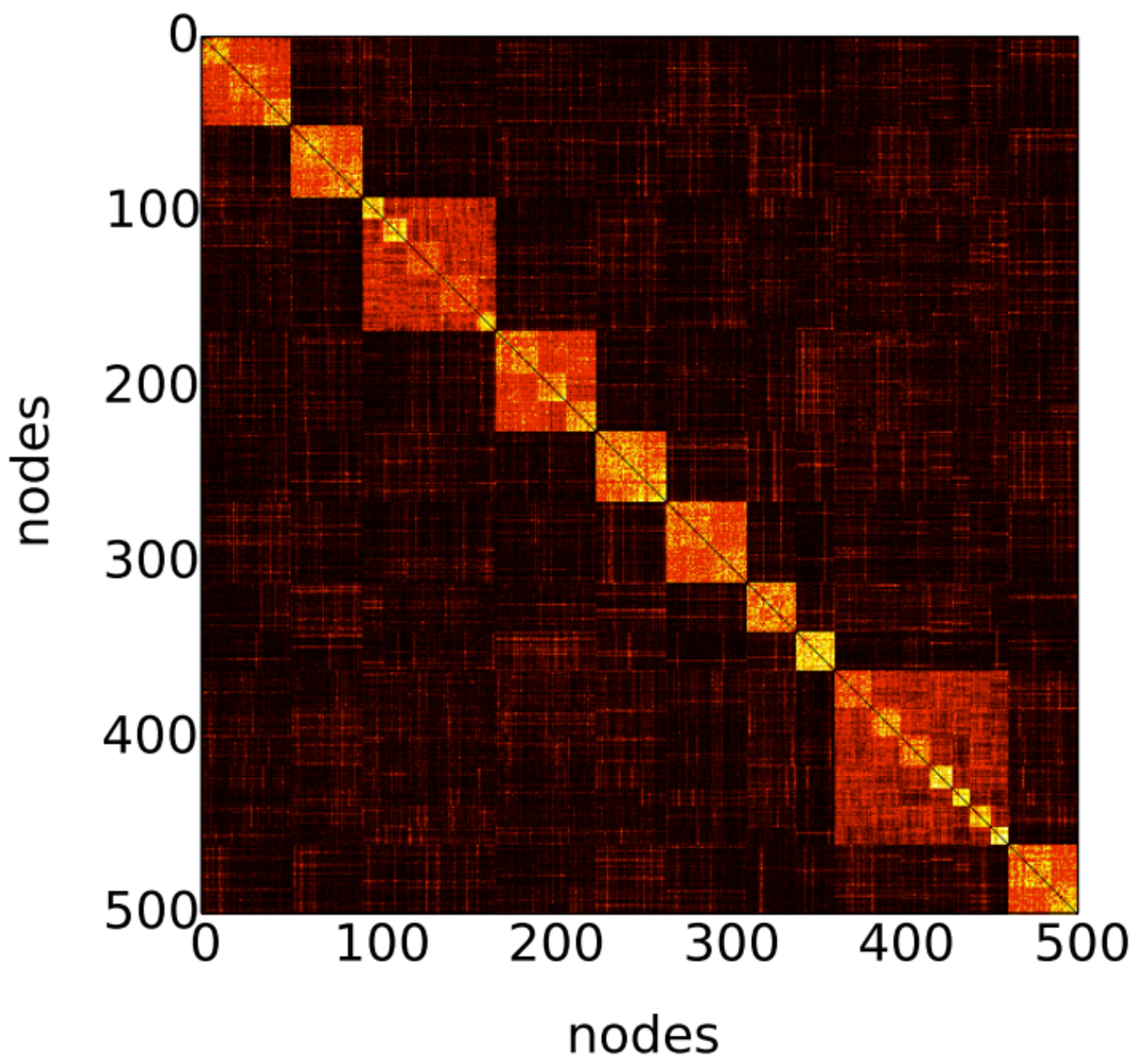}}
\caption{ (color online) Cohesion map for an undirected and unweighted benchmark network of  $N=500$ nodes and $M=7454$ edges organized into 10 first level and  28 second level communities \mbox{\cite{benchmark1,benchmark2}} (lighter squares represent higher probabilities, see colorbar in Fig.\ref{F:intro}).}
\label{F:multilevel}
\end{center}
\end{figure}

{\cz Let us consider an undirected graph with $N$ nodes (or vertices) and $M$ links (or edges). 
A link is a connection between two nodes. 
The distance between two nodes is measured as the number of edges (``hop count") along the shortest path between the respective nodes.  
As defined in \cite{graphvoronoi} for creating the graph-Voronoi diagram {\cz a number of $g$ generator nodes are chosen from the graph and all the other nodes are} associated with the ``closest'' generator node. 
If a node is at the same distance from several generators we choose randomly the cell it is associated with to assure an unbiased tessellation.} 

The goal of the study presented in \cite{Deritei14} was to delimitate {\cz communities} as graph-Voronoi cells associated with carefully chosen generator nodes [Fig. \ref{F:voronoi}(b)]. Here we abandon the procedure of identifying such centers and, instead, an ensemble of  graph-Voronoi diagrams are generated wherein each {\cz diagram} is obtained from a set of generator nodes drawn randomly from the network.

The steps of the algorithm are as follows:

\begin{itemize}

 \item[(i)] We randomly pick $g$ nodes from the network and using these as generator nodes we perform a graph-Voronoi tessellation. Subsequently, for each pair of nodes we determine whether they are co-located, i.e., share the same Voronoi cell or not  [see the twelve node network and the co-location matrix thereunder in Fig. \ref{F:intro}].

 
 \item[(ii)] We repeat the above tessellation {\cz $R$} times  and calculate the Voronoi cohesion matrix defined as the ensemble average of the co-location matrix, i.e., the co-location probabilities for all node pairs [see the matrix on the right side of Fig. \ref{F:intro}]. 
\item[(iii)] {\cz The marked block structure  of the cohesion matrix allows for the automated identification of the loosely connected clusters making up the network. Details are offered in Sec. \ref{S:boost}.}
\end{itemize}

As another example in Fig. \ref{F:multilevel} we present the Voronoi cohesion matrix obtained on a multilevel benchmark network (see the Appendix) with $N=500$ nodes,  $M=7454$ edges, 28 communities on first level and 10 communities on the second level \cite{benchmark1,benchmark2}. 
The size of generator sets, i.e., the number of generator nodes, $g=30$ in the case shown, is a fixed parameter of the ensemble.   
 {\cz For a better visualization when plotting the cohesion matrix for benchmark networks nodes are ordered according to the ground truth, or simply truth, i.e., \textit{a priori} known community information (see the Appendix), so that the matrix features a visually apparent block diagonal structure.} 
 
 {\cz It is common that for real world networks no truth information is available.  When plotting their cohesion matrix we will use the output of mainstream community detection methods  as substitute for truth information. }

\section{Analytical approach for large non-hierarchical networks }\label{S:analytical}

In this section we study analytically the stochastic  graph-Voronoi tessellation on large non-directed random networks of $N\rightarrow\infty$ nodes made up of $m$ non-overlapping but connected Erd\H{o}s-R\'enyi (ER) type modules {\cz (see the Appendix)} of size $N_i\equiv\alpha_i N,\ 0<\alpha_i<1$ and connectivity characterized by the edge density matrix  $q_{ij}=M_{ij}/N_iN_j$, $i,j\in{1,..,m}$, i.e., the probability for an edge with endpoints in modules $i$ and $j$ to exist. 
$M_{ij}$ stands for the number of edges connecting the two modules.
The diagonal elements of this matrix, also denoted as $q_i$, will be referred to as \textit{intra-module edge densities}  while off-diagonal elements as \textit{inter-module edge densities}. 

{\cz Henceforth, as a synonym of ``inter-module" we will also use the word ``bridge" and the symbol $b$ to denote the corresponding edge densities. }
By bridge nodes we mean those vertices that sit on inter-module (bridge) edges. 
The number of those in module $i$ creating a bridge to module $j$ is $B_{ij}$.  

{\cz Hereafter,} lower indices of all quantities, unless otherwise stated, refer to the corresponding modules. 
By ``cell'' we always mean graph-Voronoi cell. For more details on graph-Voronoi diagrams see \cite{Deritei14,graphvoronoi}. 
  
Let us introduce the following events:
\begin{itemize}
 \item[(i)] $X_{ij}$ - two nodes from  modules $i$ and $j$, respectively,  belong to the same cell ;
 \item[(ii)] $G_{n_1n_2\dots n_m},\ \sum_{i=1}^m n_i = g$ - the $g$ generator nodes are shared among the $m$ modules so that $n_i$ generators fall into module $i$. 
\end{itemize}
$\{G_{n_1n_2\dots n_m}\}$ is a complete set of $C_{g+m-1}^{m-1}$ events (number of realizations for segmenting a linear chain of $g$ identical 
balls by $m-1$ randomly placed walls. $C_n^k$ denotes the binomial coefficient for the $k$-combinations of $n$ elements). Therefore
\be
 X_{ij} = \sum_{{n_1n_2\dots n_g}} X_{ij} \cdot G_{n_1n_2\dots n_g}\ .
\ee
The Voronoi cohesion, i.e., the probability for two nodes from modules $i$ and $j$, respectively, to belong to the same cell is
\[
 c_{ij}\equiv P(X_{ij}) = \sum_{{n_1n_2\dots n_g}}P(X_{ij}|G_{n_1n_2\dots n_g})P(G_{n_1n_2\dots n_g})\ ,
\]
or in a more compact form
\be\label{masterEq}
\vec{c} = \vec{V}\cdot\vec{g}\ ,
\ee
where $\vec{c}, \vec{V}$ and $\vec{g}$ are matrices of size $N(N+1)/2$-by-1, $N(N+1)/2$-by-$C_{g+m-1}^{m-1}$ and $C_{g+m-1}^{m-1}$-by-1, respectively. 

In order to get a better grasp of the employed analytical tools we shall  consider two special cases where cohesion, contrast and other statistical measures can be estimated.

\subsection{Case 1: Extreme modularity}\label{S:analyticalextreme}

In this setup we assume the number of inter-module (bridge) edges, $M_{ij}$, {\cz $i\neq j$}, to be of  {\cz order} $\mathcal{O}(N^{\beta}), 0 < \beta < 1$, i.e., negligible compared to that of intra-module edges, $M_{ii}$, and equal to the number $B_{ij}$ of the bridge nodes. 
Consequently, in the $N\rightarrow\infty$ limit the ratio  $B_{ij}/N_i$ of bridge nodes will also be negligible and their inter-module degree will be one.

Let us consider a \textit{two module network} with \textit{two generator nodes}. 
The modules, generators and their respective Voronoi cells will be labeled and referred to using the numbers 1 and 2. 
The disparity between intra- and intermodular cohesion values can be characterized by the \textit{contrast} defined as
\be
\gamma\equiv\frac{c_{11}+c_{22}-2c_{12}}{c_{11}+c_{22}+2c_{12}}\ , \qquad -1\leq\gamma\leq 1\ \text{when}\ 0\leq c_{ij}\leq 1\ .
\label{contrast2}
\ee
In the limit of a single homogeneous network ($c_{11}=c_{22}=c_{12}$) the communities are not discernible  and the contrast vanishes. 
The larger its value the easier to identify the clusters.
Ideally, a disconnected two-module network should yield very low intermodular cohesion ($c_{12}\ll 1$) corresponding to a contrast close to unity. 
 As shown in \cite{Blondel07} for an infinitely large ER network with link density $q$, the probability distribution, $f$, of the inter-node distance $d$ is: $f(d=1)=q, f(d=2) =  1-q$ and $f(d \geq 3)=0$.
Non-bridge nodes will be directly connected to an average of $q_iB_{ij}$ bridge nodes and the probability $(1 - q_i)^{B_{ij}}$ for lacking any direct links to the bridge vanishes.  
Thus, apart from the infinitesimal bridge, any two nodes belonging to different modules are three steps away. 
Therefore the probability distribution of distances, $f_{ij}(d)$, between two nodes from modules $i$ and $j$, respectively, takes the following values:
\begin{center}
\begin{table}[!htbp]
\begin{ruledtabular}
\begin{tabular}{c|cccccc}
\backslashbox{$ij$}{$d$}  & 1     &       2 & 3 \\\hline
1,1                       & $q_1$ & $1-q_1$ & 0 \\
1,2                       &     0 &       0 & 1 \\
2,2                       & $q_2$ & $1-q_2$ & 0 
\end{tabular}
\end{ruledtabular}
\end{table}
\end{center}
Consequently, if the generator nodes are picked from different modules they will both ``expropriate" the whole respective module. However, when they belong to the same module both modules will be shared equally between the two Voronoi cells.
Synthesizing the probabilities for the different configurations and applying (\ref{masterEq}) we get: 
\bgroup
\def\arraystretch{1.5}
\begin{center}
\begin{table}[!htbp]
\begin{ruledtabular}
\begin{tabular}{c|c|ccccc|c|ccc|c|c}
\multicolumn{2}{c|}{\multirow{2}{*}{$\vec{V}$}} & 
\multicolumn{3}{c}{$n_1n_2$} & 
\multirow{2}{*}{}& 
\multicolumn{2}{c|}{\multirow{2}{*}{$\vec{g}$}} & &
\multirow{2}{*}{}& 
\multicolumn{2}{c|}{\multirow{2}{*}{$\vec{c}$}} &
\\\cline{3-5}
\multicolumn{2}{c|}{}  &    2,0 & 1,1  & 0,2  &  & \multicolumn{2}{c|}{} & & & \multicolumn{2}{c|}{} & \\
\cline{1-5}\cline{7-9} \cline{11-13}
\parbox[t]{3mm}{\multirow{3}{*}{\rotatebox[origin=c]{90}{$i,j$}}}  & 
1,1 & 1/2 & 1 & 1/2& &
\parbox[t]{2mm}{\multirow{3}{*}{\rotatebox[origin=c]{90}{$n_1n_2$}}}& 2,0 & $\alpha_1^2$
& &\parbox[t]{3mm}{\multirow{3}{*}{\rotatebox[origin=c]{90}{$i,j$}}}  & 1,1 &$1/2+\alpha_1\alpha_2$\\
& 1,2   & 1/2 & 0 & 1/2 & &  & 1,1 &$2\alpha_1\alpha_2$& & & 1,2   &$1/2-\alpha_1\alpha_2$\\
& 2,2   & 1/2 & 1 & 1/2 & &  & 0,2 &  $\alpha_2^2$ & &  & 2,2   &$1/2+\alpha_1\alpha_2$   
\end{tabular}
\end{ruledtabular}
\end{table}
\end{center}

It is remarkable that cohesion does not depend on the edge density and is uniform across the modules, i.e. $c_{11}=c_{22}$, 
even for very unbalanced module size distribution. The contrast defined in (\ref{contrast2}) becomes
\be\label{nobridge_contrast_limit}
\gamma(\delta)=2\alpha_1\alpha_2=\frac{1}{2} - 2\delta^2\ ,
\ee
where $\delta \equiv \alpha_1 - 1/2 = 1/2 - \alpha_2 \in (0, 1/2)$ characterizes the  imbalance in the size of the modules. 
The above defined contrast is maximal for equally large modules. 
In the context of extreme modularity this value,  $\gamma(0) = 1/2$, is counterintuitively low as near unity values are rather expected. 
The discrepancy can be attributed to the fact that in half of the cases the generators fall into the same module and both modules will be shared evenly between the two cells. 

This effect is mitigated when considering \textit{two modules and a number of $g$ generators}. 
Here any module is shared evenly either among the $k\geq 1$ generators placed in the respective module or, if $k=0$, among all $g$ generators. The ingredients of the master equation (\ref{masterEq}) are

\be\label{masterEqTwoModules}
\vec{V} = \left(\begin{array}{ccccccc}
1/g & 1 & 1/2 & \cdots & 1/k & \cdots & 1/g\\
1/g & 0 & 0 & \cdots & 0 & \cdots & 1/g  \\
1/g & 1 & 1/2 & \cdots & 1/k & \cdots & 1/g
\end{array}\right)\ ,
\ee
\begin{align*}
\vec{g} =(B_{\alpha_1}^g(0), B_{\alpha_1}^g(1),\cdots,B_{\alpha_1}^g(k), \cdots,B_{\alpha_1}^g(g))^\intercal\ ,\\
\quad B_{q}^g(k) = C_g^kq^k(1-q)^{g-k}\ ,\quad \vec{c} = (c_{11}, c_{12}, c_{22})\ ,
\end{align*}
whence the relevant cohesion values become
\be\label{cohTwoModuleMultiGenerator}
 c_{11}=\frac{1}{g}(\alpha_1^g + \alpha_2^g) + \sum_{k=1}^{g-1} \frac{1}{k} C_g^k\alpha_1^k\alpha_2^{g-k}\ ,\qquad c_{12} = \frac{1}{g}(\alpha_1^g + \alpha_2^g)\ .
\ee 
The difference in the intra-module cohesions is:
\begin{align*}
 c_{11} - c_{22} =& \sum_{k=1}^{g-1} \left(\frac{1}{k} - \frac{1}{g-k}\right)C_g^k\alpha_1^k\alpha_2^{g-k}=\\
 =&8g(\alpha_1\alpha_2)^{g/2}\sum_{l=1-g/2}^{g/2-1} \frac{l}{4l^2 - g^2}C_g^{l+g/2}\left(\frac{\alpha_1}{\alpha_2}\right)^l\ .
\end{align*}
The summation interval is symmetric to zero, for $g > 2$ the first factor is an odd function of  $l$, while the binomial factor is only even
when $\alpha_1=\alpha_2=1/2$. Therefore, $c_{11} >(<) c_{22}$ when $\alpha_1 > (<) 1/2$. 
\begin{figure}[htbp]
 \begin{center}
  \resizebox{1.0\columnwidth}{!}
{\includegraphics{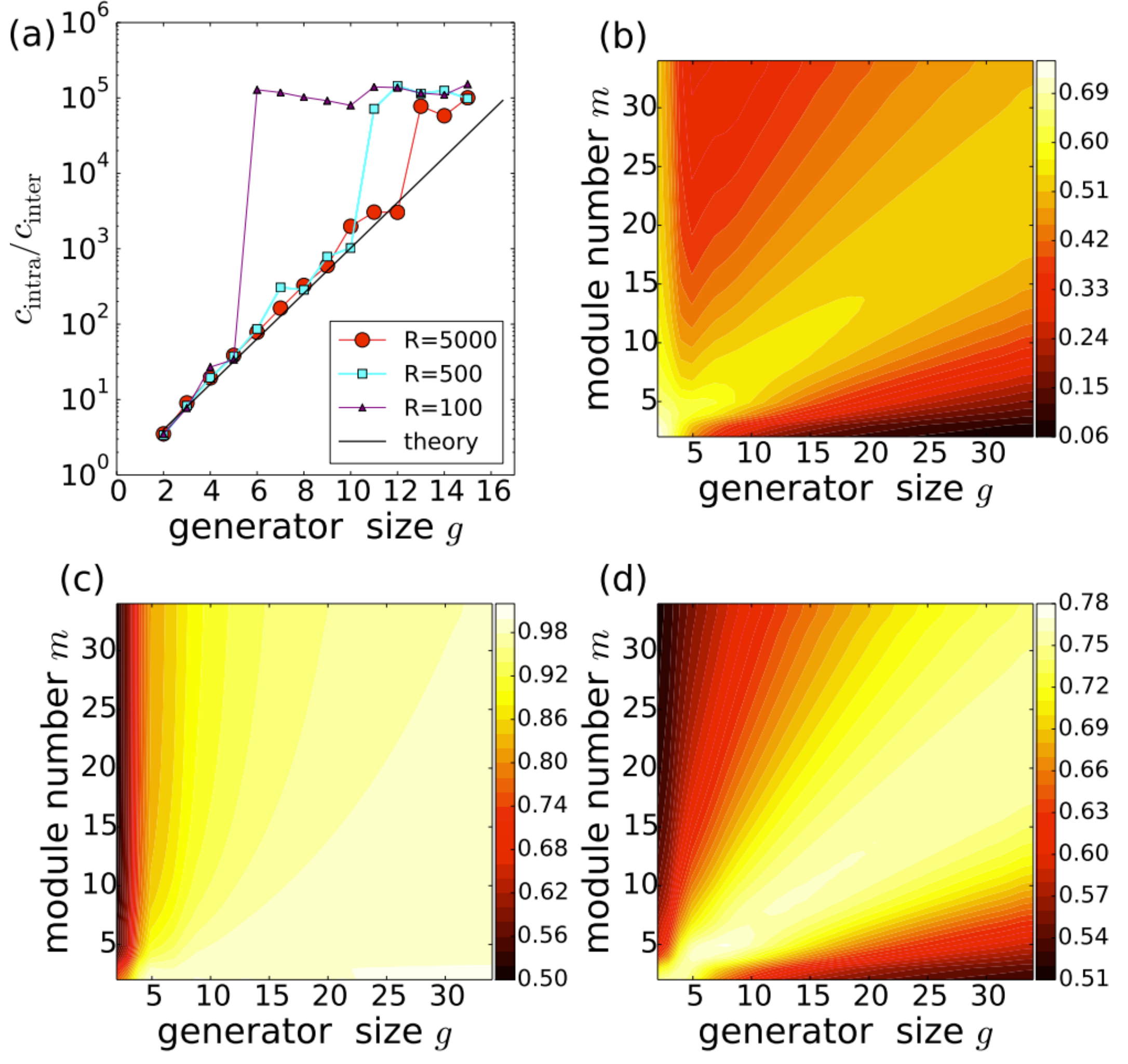}}
 \caption{ (color online) Dependence of intra- and intermodular cohesion ratio $c_{11}/c_{12}$ on the number of generator nodes for the case of a two-module network of  extreme modularity. 
 (a) Simulation results for an extremely modular network of size $N=2\times 250$ with $m=2$ ER-type modules, with intra- and inter-module link densities of $q=0.5$ and $b=10^{-4}$, respectively. 
 Voronoi cohesion matrices were estimated for three different ensemble sizes (repeats) ($R=100, 500$ and 5000).  The analytical results expressed by Eqs. (\ref{extremeModularityIntraMEq2}) and (\ref{extremeModularityInterMEq2}) are represented as continuous black line. (b), (c), (d) Dependence of sensitivity [$c_\text{intra}$ from Eq. (\ref{sensitivity})], specificity [$1-c_\text{inter}$ from Eq. (\ref{specificity})] and accuracy [$(c_\text{intra} - c_\text{inter} + 1)/2$] on generator size and module number \cite{statistics}. Lighter colors represent higher values.
 }
 \label{F:extremeModularityGeneratorDependence} 
\end{center}
\end{figure}

For understanding the generic dependence of the cohesion based separability on the number of generators we should first consider the instructive example of the two extremes: a single generator ($g=1$) forges the whole network into the same cell; and, conversely, the maximum number of generators ($g=N$) associates all nodes with a different cell. 
In terms of statistical classification functions \cite{statistics} the sensitivity, i.e., proportion of positives (pairs in the same cell) that are correctly identified as such (belong to the same module), also known as true positive rate (TP), and specificity, i.e., the proportion of negatives (pairs in different cells) that are correctly identified as such (belong to different modules) or true negative rate (TN) mutually exclude each other in the above diametric cases  (for $g=1, TP=1, TN=0$; for $g=N, TP=0, TN=1$). 
Clarifying the apparent trade-off requires a more quantitative approach. 

Confining the study to the case of $m$ equally large modules connected to one another in a uniform fashion is sufficient to form a general picture. 
In this simplified setup with equivalent modules a node pair either belongs to the same or to different modules. 
The cohesion for the intramodular case is identical to sensitivity (true positive rate):
\begin{align}
TP&\equiv c_\text{intra} = P(X_{ii})= \sum_{k=0}^g P(X_{ii}|n_i=k)P(n_i=k)=\nonumber\\
  &=\frac{1}{g}B_{1/m}^g(0) + \sum_{k=1}^{g}\frac{1}{k}B_{1/m}^g(k)\ ,\qquad\qquad \forall i\in{1,..,m}\label{sensitivity}
\end{align}
where $n_i$ is the number of generators in module $i$. 
For $m=2$ and $g\gg 1$
\be\label{extremeModularityIntraMEq2}
c_\text{intra} = \dfrac{2^{-g}}{g}+2^{-g}\sum_{k=1}^g C_g^k\frac{1}{k}
\approx \dfrac{2^{-g}}{g} + \frac{2}{g}=\mathcal{O}\left(\frac{1}{g}\right)\ .
\ee
Above we made use of the fact that the binomial coefficient, $C_g^k$, {\cz when represented graphically as a function of $k$} is a bell-shaped curve centered on $g/2$ with a relative spread decaying as $1/\sqrt{g}$. 

If $m \gg g$ 
\[
c_\text{intra}\approx \frac{1}{g} + \frac{g-1}{m}\ ,
\]
while for $m,g\gg 1$ and $m=g/\lambda$ the binomial distribution is well approximated by a Poissonian yielding
\be\label{extremeModularityIntraLambda}
c_\text{intra}\approx \frac{e^{-\lambda}}{g}+f(\lambda)=f(\lambda) + \mathcal{O}\left(\frac{1}{g}\right)\ ,
\ee
where
\[
f(\lambda)=e^{-\lambda}\sum_{k=1}^\infty \dfrac{\lambda^k}{k!}\frac{1}{k} =-e^{-\lambda}\left[\gamma_e + \Gamma(0, -\lambda) + \log(-\lambda) \right]\ ,
\]
with $\gamma_e=0.5772156...$ as the Euler-Mascheroni constant and $\Gamma(s,x) = \int_x^{\infty} t^{s-1}\,e^{-t}\,{\rm d}t$ the upper incomplete gamma function. $f(\lambda)$ reaches its $\approx 0.52$ peak value at $\lambda\approx 1.5$ and decays as $1/\lambda$ for $\lambda \gg 1$. The above formulas indicate that the large number of modules is detrimental to sensitivity. As also revealed by Fig. \ref{F:extremeModularityGeneratorDependence}(b) this can be partially overcome by  \textit{choosing a generator size approximately 50\% higher than the number of modules}. Later sections will not only validate this finding but will also invest it with practical relevance.

The probability that two nodes chosen from different modules are co-located (false positive) is $1/g$ if at least one of the two modules is seedless: 
\begin{align}
FP&\equiv c_\text{inter}=P(X_{ij})=\qquad\qquad\quad \forall i\neq j\in{1,..,m}\nonumber\\
  &=\frac{1}{g} \left[P(n_i = 0) + P(n_j = 0) - P(n_i=n_j=0)\right]=\nonumber\\
  &= \left.\left[2B_{1/m}^g(0)-B_{2/m}^g(0)\right]\middle/ g\right.=\nonumber\\
  &=\frac{1}{g}\left[2\left(1-\frac{1}{m}\right)^g - \left(1-\frac{2}{m}\right)^g\right]\ .\label{specificity}
  \end{align}

For the asymptotic behaviors of interest we get  
\begin{align}
\frac{2}{g}2^{-g} \ , \qquad &\text{for}\ m=2\ ,\label{extremeModularityInterMEq2}\\
\frac{1}{g} - \frac{g-1}{m^2}\ , \qquad &\text{for}\ m\gg g\ ,\nonumber\\
\frac{e^{-\lambda}}{g}\left[2 - e^{-\lambda}\right]\ , \qquad &\text{for}\ m,g\gg 1\ ,\ \text{and}\  m=g/\lambda\ .\label{extremeModularityInterLambda}
\end{align}

Figure \ref{F:extremeModularityGeneratorDependence}(a) illustrates the verification by simulation of the $m=2$ case for the $c_\text{intra}/c_\text{inter}$ quantity exhibiting a $\mathcal{O}\left(2^g\right)$ dependence on the generator size. The deviation from the theoretical curve emerging from Eqs. (\ref{extremeModularityIntraMEq2}) and (\ref{extremeModularityInterMEq2}) is due to the finite ensemble size, $R$, limiting the accuracy of the cohesion estimation.

Specificity can be directly related to $c_\text{inter}$ by the formula $TN = 1 - FP$. Its dependence on the number of modules and generators is illustrated in Fig. \ref{F:extremeModularityGeneratorDependence}(c). Instead of the contrast defined in Eq. (\ref{contrast2}) a more common way for characterizing the quality of the co-location detection is accuracy, defined as the arithmetic mean of the sensitivity and specificity, i.e., $(c_\text{intra} - c_\text{inter} + 1)/2$. On Fig. \ref{F:extremeModularityGeneratorDependence}(d) the linear shape of the level curves can be understood from the asymptotic limits described by Eqs. (\ref{extremeModularityIntraLambda}) and (\ref{extremeModularityInterLambda}). 

Accuracy, especially within the framework of the above oversimplified model, might not be the most suitable measure for appreciating the practical relevance of the obtained cohesion picture. Nevertheless, we expect the main messages conveyed by this section to be applicable in a much wider context. 

\subsection{Case 2:  Non-infinitesimal intermodular connection density}\label{S:analyticalfinite}
\begin{figure}[!htbp]
 \begin{center}
  \resizebox{1.0\columnwidth}{!}
 {\includegraphics{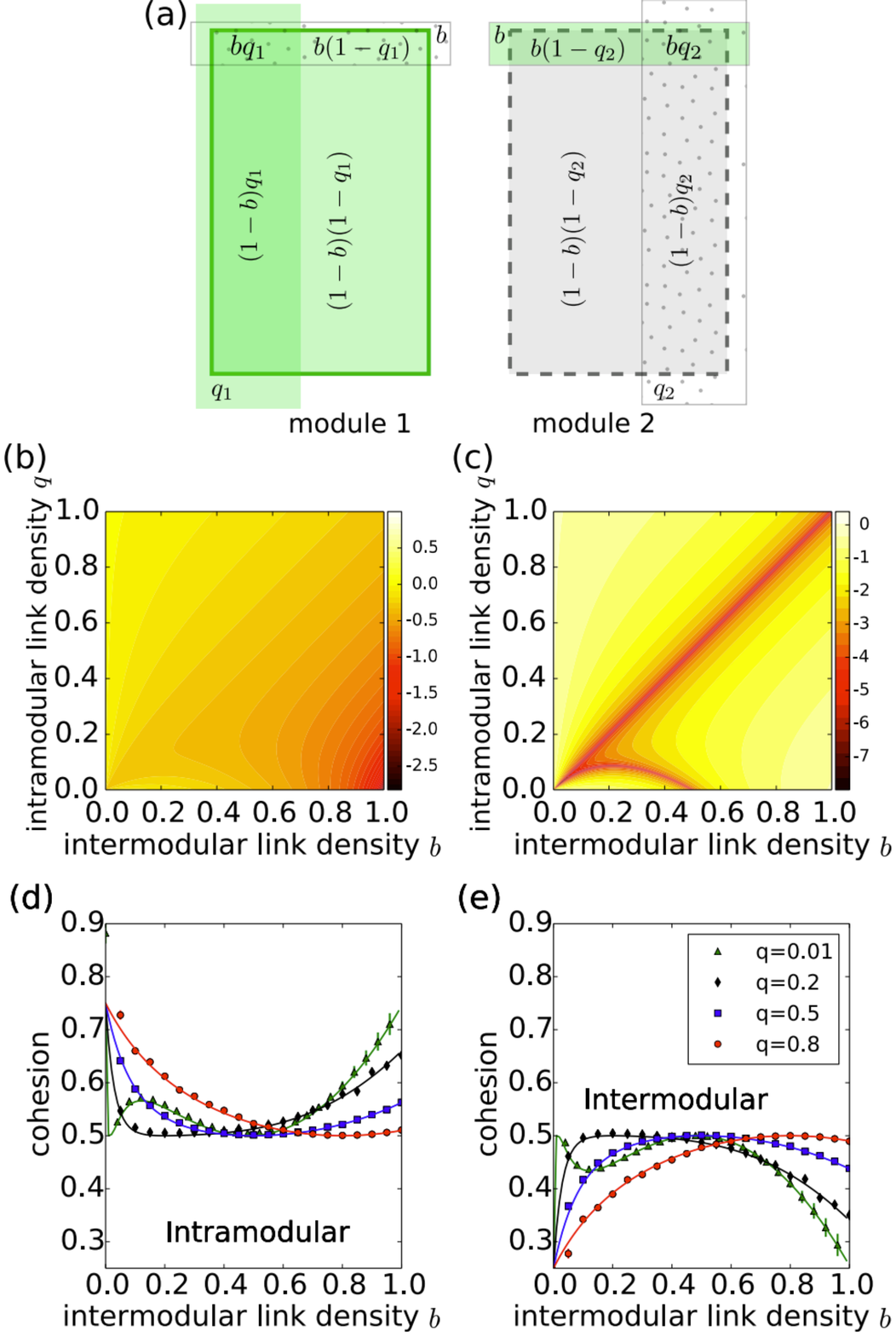}}
 \caption{ (color online)  Study of Voronoi cohesion for a network of two equally large ER-type modules ($m=2$) with non-infinitesimal intermodular link density, $b$. There is one generator placed into each of the modules ($g=2$). 
 (a) Illustration of how the two modules are shared between the two Voronoi cells  as a function of the distance to the generators. The link densities in the two modules are $q_1$ and $q_2$, respectively. Areas covered by flatly colored and dotted secondary rectangles represent ``regions'' with nodes at unit distance  (one step away) from generator one and two, respectively. 
 Remaining areas represent nodes that are at distance two from both generators. 
 (b) Influence of the intra-, $q_1=q_2=q$, and intermodular densities, $b$, on the relative ``area" of cell 1 in module 1  [see Eq. (\ref{bq_s})]. 
 (c) The corresponding contrast calculated from Eq. (\ref{contrast_non_extreme_modularity}). 
 (d), (e)  Intra- and intermodular Voronoi cohesion as a function of the intermodular link density, $b$. Analytical results expressed in Eqs. (\ref{bq_c11}) and (\ref{bq_c12}) are compared to simulation for several intramodular link densities, $q$. Simulation data was obtained from an ensemble of 5000 tessellations (10 topologies $\times$ 500 generator sets) of a network of $N=2\times800$ nodes. Most error bars representing standard error lay within the markers.}\label{F:non_extreme} 
 \end{center}
\end{figure}
Once the extreme modularity requirement is relaxed the density of the inter-module connections acquires primary focus. Let us confine the investigation to the case of two similar modules, both of size $N\gg 1$ and link density $q$. 
The intermodular connection (bridge) density $b$ denotes the probability for any given inter-module link to exist. {\cz The number of both intra- and inter-module links is of order $\mathcal{O}(N^2)$}.

Below we summarized some relevant quantities of a pair of nodes as a function of their location (in module 1. or 2.). 
The left table contains the probability distribution of  their mutual distances. Assuming that they are both generators  the right table collects the relative sizes of the corresponding Voronoi cells in each of the modules: 
\begin{center}
\begin{table}[!htbp]
\begin{ruledtabular}
\begin{tabular}{c|ccccc|cc|c}
\multirow{2}{*}{\parbox[t]{8mm}{\centering loc. in mod.} }  & 
\multirow{2}{*}{\rotatebox[origin=r]{90}{$d=1$}} &  \multirow{2}{*}{\rotatebox[origin=r]{90}{$d=2$}} & 
\multirow{2}{*}{} & 
\multicolumn{2}{c|}{module 1} & \multicolumn{2}{c|}{module 2} & \multirow{2}{*}{\parbox[t]{8mm}{\centering loc. in mod.}} \\\cline{5-8}
& & & & gen.1     &    gen.2    &    gen.1    &     gen.2     &                        \\\cline{1-3}\cline{5-9}
1,1                                        &        $q$         &       $1-q$   &
& 1/2      &      1/2     &     1/2      &      1/2       &           1,1       \\
1,2                                        &        $b$         &       $1-b$    &
&  $s$      &     $1-s$    &    $1-s$     &      $s$       &           1,2          \\
2,2                                        &        $q$         &       $1-q$     &
&  1/2      &      1/2     &     1/2      &      1/2       &           2,2 
\end{tabular}
\end{ruledtabular}
\end{table}
\end{center}
where  $s$ is the relative size of cell 1 in module 1 when generator  2 is placed in module 2. Its value can be obtained by combining the following contributions [see Fig. \ref{F:non_extreme}(a)].
\begin{itemize}
 \item[(i)] $bq$ ratio of the module 1. is directly connected to both generators and is equally shared between the two cells.
 \item[(ii)] $q(1-b)$  is directly connected to generator 1 only; therefore, it contributes in full to cell 1.
 \item[(iii)] $(1-q)b$ is directly connected only to generator 2 therefore does not contribute to cell 1.
 \item[(iv)] $(1-b)(1-q)$ is indirectly connected to both generators. 
 It will be split proportionally to the number of shortest paths to the two generators. 
 The ``stepping stone'' node to generator 1 can be from both modules with a probabilities of $q^2$ and $b^2$, respectively. 
 The similar quantities for the ``remote'' generator 2 are both $bq$. Thus a ratio of $(q^2 + b^2)/(q+b)^2$ of this domain goes to cell 1. 
\end{itemize}
Summing up the different contributions we get:
\be\label{bq_s}
 s(q,b) = \frac{qb}{2} + \dfrac{1-b}{(q+b)^2}\left[q^2 + b^2 + 2q^2b\right]\ .
\ee
Figures \ref{F:non_extreme}(b) and \ref{F:non_extreme}(c) show the dependence of this size on the intra- and inter-community densities and the associated 
contrast  exhibiting the expected limiting behavior, e.g., $s(q, 0) = 1, s(q, q) = 1/2$.

The master equation (\ref{masterEq}) can now be constructed from the matrices:
\[
\vec{V} = \left(
\begin{array}{ccccccc}
 1/2 & 1-2s(1-s) &1/2\\
 1/2 & 2s(1-s)   &1/2\\
 1/2 & 1-2s(1-s) &1/2
\end{array}\right)\ ,\quad 
\vec{g} =\left(
\begin{array}{c}
1/4\\ 1/2\\ 1/4
\end{array}
\right)\ .
\]
Therefore
\be\label{bq_c11}
 c_{11} = c_{22} = \frac{3}{4} - s(1-s)\ ,
\ee
\be\label{bq_c12}
 c_{12} = \frac{1}{4} + s(1-s)\ .
\ee
Hence the link density dependent contrast is
\be\label{contrast_non_extreme_modularity}
 \gamma=\frac{1}{2}\left[1-4s(1-s)\right]\ . 
\ee
If $b=q$, that is, the inter- and intramodular link densities coincide, $\gamma=0$, meaning that the network ceases to be modular. 
Similarly to Eq. (\ref{nobridge_contrast_limit}) the maximum contrast, $\gamma=1/2$, can be obtained when the bridge density, $b$, is negligible. 

Simulations aptly confirm the analytical results.  
Fig. \ref{F:non_extreme}(d) and \ref{F:non_extreme}(e) show the cohesion values obtained for a network built from two modules of $800$ nodes each connected at different inter- and intramodular link densities. 
For each $q, b$ link density pair the corresponding cohesion matrix was obtained as the co-location grand mean of an ensemble of ten random topologies each subjected to 500 random tessellations. 
Average cohesion follows closely the theoretical curve from Eqs. (\ref{bq_c11}) and (\ref{bq_c12}). 

One striking property of the method is that even a relatively small ensemble and network size is sufficient to confirm theory. 
This aspect facilitates to a great extent the adoption of stochastic graph Voronoi diagrams for community detection purposes. 

\section{Complexity vs. accuracy}\label{S:complexity}
\begin{figure*}[htbp]
   \begin{center}
   \resizebox{0.9\textwidth}{!}
   {\includegraphics{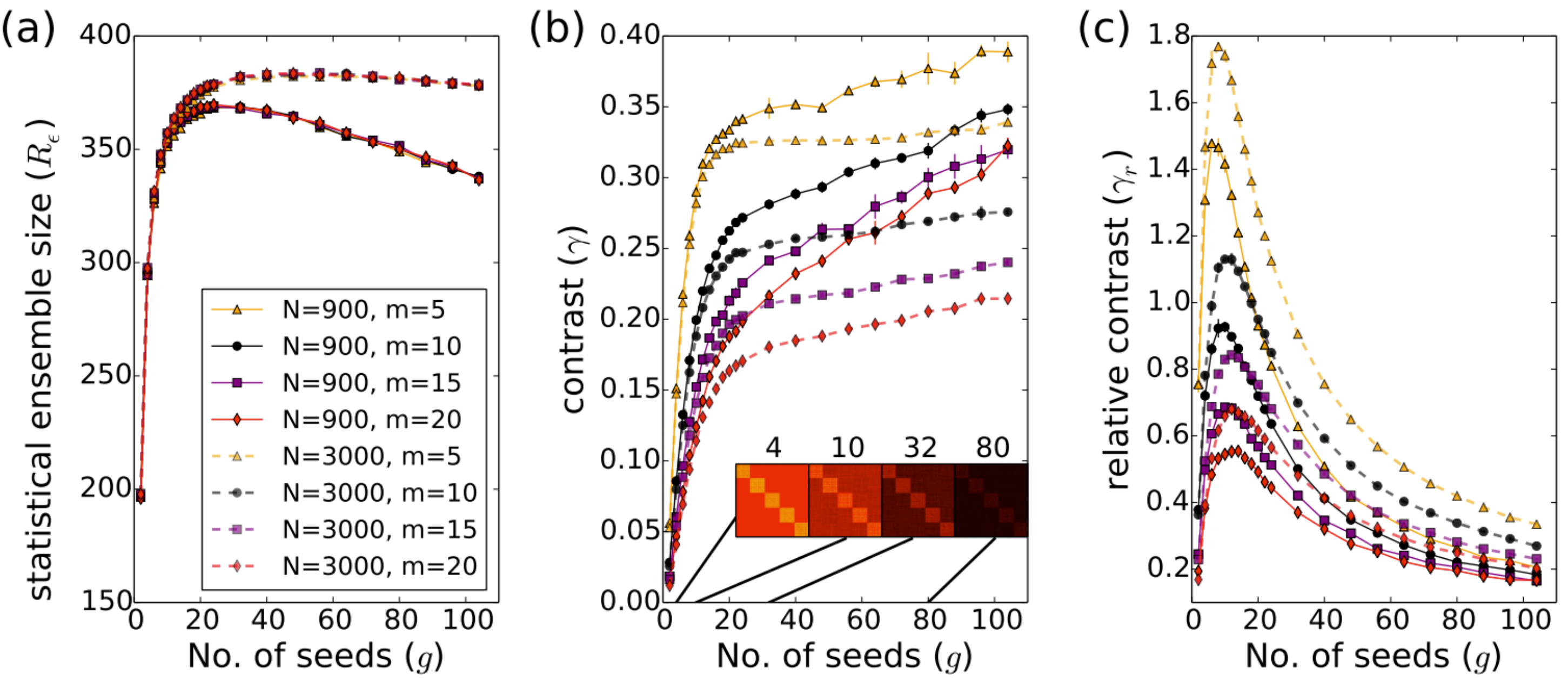}}
    \caption{ (color online)  Simulation study of the effect of network size, $N$, number of ER-type modules, $m$, and the number of seeds (generators), $g$. Intra- and intermodular link densities are set to $q$=0.5 and $b$=0.05, respectively. (a)  Statistical ensemble size (number of iterations) necessary to keep the estimation error of the cohesion below a fixed tolerance threshold {\cz of $5\times 10^{-4}$}[see Eq. (\ref{error_iter})]. (b) Contrast defined in Eq. (\ref{contrast3}). The ruined cohesion picture (see insets) at large seed numbers confirms the conclusions of Sec. \ref{S:analyticalextreme} and demonstrates the unsuitability of contrast as defined in Eqs. (\ref{contrast2}) and (\ref{contrast3}) for situations when Voronoi cells are small. (c) Relative contrast defined in Eq. (\ref{contrast_relative}).}
    \label{F:effect_of_seed}
    \end{center}
\end{figure*} 
Our concern is the computational costs incurred by producing the Voronoi cohesion matrix. We are primarily interested in the impact of the number of seeds and of basic network properties such as network size and number of modules. In order to surmount the limitations imposed by the idealizations from the previous section we shall formulate our conclusions based on simulations. 

The complexity of an ensemble of graph-Voronoi partitioning is $O(R N\log{N})$ \cite{graphvoronoi} where {\cz $R$} is the size of the ensemble, i.e., the number of generated graph-Voronoi diagrams. 
As stated earlier the Voronoi cohesion matrix is the expectation value of the tessellations' co-location matrix  wherein unit value represents node pairs sharing the same cell and zero otherwise. 
The ratio converges toward the ``true" cohesion in the $R \rightarrow \infty$ limit. 
 
For practical purposes the size of the ensemble should be a function of the acceptable statistical error, $\epsilon$, hereinafter, tolerance. Henceforth, instead of ensemble size we shall mostly refer to its inverse, $1/R$, under the name \textit{convergence rate}.
The cohesion matrix is the limit of the sequence
\begin{equation}\label{coh_iter}
 \vec{C}_{n+1} = \frac{n\ \vec{C}_{n} + \vec{C}}{n+1}\ ,\qquad 0 < n < R\ 
\end{equation}
where $\vec{C}$ is the co-location matrix in the ($n+1$)st step, while the error is estimated as
\begin{equation}\label{error_iter}
 e_{n+1} = \frac{\| \vec{C}_{n+1} - \vec{C}_{n} \|}{ \| \vec{C}_{n}\|} = \frac{\| \vec{C} - \vec{C}_{n} \|}{(n+1) \| \vec{C}_{n}\|} > \epsilon\ ,
\end{equation}
where the norm $\|\cdot\|$ is based on some arbitrary metric normalized to the number of dimensions, in this case $N^2$. 

The study of complexity comes down to the estimation of the size of the ensemble to be generated for a given tolerance. 
However, the convergence rate of the iteration in Eq. (\ref{coh_iter}) might be influenced by factors such as the number of generators, 
number of modules and the size of the network. A straightforward generalization of the contrast in Eq. (\ref{contrast2}) for multimodular networks could be
\begin{equation}\label{contrast3}
\gamma = \dfrac{\overline{c}_{\text{intra}} - \overline{c}_{\text{inter}}} {\overline{c}_{\text{intra}} + \overline{c}_{\text{inter}}}\ ,
\end{equation}
where 
\[
\overline{c}_{\text{intra}} \equiv \dfrac{\sum_{i\neq j} C_{ij}T_{ij}}{n_{\text{intra}}} \ ,\qquad
{n}_{\text{intra}} \equiv \sum_{i\neq j} T_{ij}\ ,
\]
and $C_{ij}, T_{ij}$ are the off-diagonal elements of the cohesion matrix, $\vec{C}$, and truth matrix, $\vec{T}$. The latter contains ones, for node pairs sharing the same module, and zeros, otherwise. 
Summation indices $i$ and $j$ run over all nodes from 1 to $N$. 
$\overline{c}_{\text{inter}}$ and ${n}_{\textsl{}\text{inter}}$ are obtained similarly by replacing $T_{ij}$ with $1-T_{ij}$. 

The procedure was applied on networks of various sizes and built from different number of ER-type modules.      
Figure \ref{F:effect_of_seed}(a) demonstrates that the convergence rate is sensitive neither to the size of the network nor to the number of modules. 
On the other hand it collapses as the number of seeds grows and appears to reach a minimum  followed by an ever improving convergence rate as the generator size increases. {\cz For a tolerance of $\epsilon=5\times 10^{-4}$  the corresponding statistical ensemble size, $R_\epsilon$, stays in the order of hundreds ($\sim$375) and changes by a few percent as network size grows from 900 to 3000.}
The contrast, as shown in Fig. \ref{F:effect_of_seed}(b), also seems to benefit from larger seed numbers. 
However, a closer look at the cohesion matrices [insets in Fig. \ref{F:effect_of_seed}(b))] rules this behavior an artifact as the cohesion picture of the clusters is ruined for larger seed  numbers. This behavior indicates that the contrast defined in Eq. (\ref{contrast2}) is not a measure that can live up to its name once the idealized conditions -- number of modules comparable to that of seeds -- cease to be maintained. The separation in the average cohesion of the different areas of the cohesion matrix will be irrelevant once  the Voronoi cells' size gets small and node to node variations will dominate the general picture. 
Therefore relating the larger scale, i.e., module level, differences to the low scale, i.e., node pair level, fluctuations in the cohesion will provide a measure that is more in line with the visual impression made by the cohesion matrix. 
To this end we introduce the \textit{relative contrast} and define it as

\begin{equation}\label{contrast_relative}
\gamma_r = \dfrac{\overline{c}_{\text{intra}} - \overline{c}_{\text{inter}}}{\overline{\sigma}_{\text{intra}} + \overline{\sigma}_{\text{inter}}}\ ,
\end{equation}
where
\[
\overline{\sigma}^2_{\text{intra}} \equiv 
\dfrac{
 \sum_{i\neq j}
 \left(C_{ij}T_{ij}-\overline{c}_{\text{intra}}\right)^2
}{
 n_{\text{intra}}
} \ .
\]

Figure \ref{F:effect_of_seed}(c) tells us that relative contrast has a preferred number of modules of around ten irrespective of the network size or number of modules. However, large networks with a few modules favor the formation of high contrast cohesion pictures somewhat justifying the elements of the theoretical setup from Sec. \ref{S:analytical}. A major trade-off between optimizing for convergence and contrast is apparent when comparing Figs. \ref{F:effect_of_seed}(a) and \ref{F:effect_of_seed}(c), {\cz namely, for optimal relative contrast larger statistical ensemble is required. However, this value is not sensitive to the size of the network and is completely unaffected by the number modules.}
\section{Community detection by contrast boosting}\label{S:boost}
After achieving efficient convergence, the Voronoi cohesion matrix can be used to extract the community structure of the network.  
{\cz We outline the  procedure through the example of a benchmark network with $N=500$ vertices, $M=5000$ edges organized into $m=9$ communities \cite{benchmark1}. In Fig. \ref{F:boost}(a) we plot the distribution of the Voronoi cohesion values obtained from an ensemble of $R=3000$ diagrams with $g=15$ generators. The latter value is in line with the $g\approx 1.5 m$ golden rule on the optimal generator-module ratio established in Sec. \ref{S:analyticalextreme}. The number of repeats, $R$, was set such that the statistical error in the cohesion values became negligible.}
{\cz Notice that, }the matrix of inter-node distances shows a similar pattern as the cohesion map yet with a significantly lower contrast and limited to integer numbers in the range between zero and five [see insets in Fig. \ref{F:boost}(a)]. 
Intra- and inter-module node pairs were separately tallied based on truth information (see the Appendix). 
{\cz For a ``quick and dirty'' identification of the communities  we observe that intra-module node pairs are well separated from inter-module pairs by a pronounced gap in the cohesion space [see the histogram in Fig. \ref{F:boost}(a)]. 
The simplest way to convert the cohesion matrix into information on community structure  is by applying a \textit{cohesion threshold} to the histogram around the minimum point of this gap.
Subsequently communities can be circumscribed as follows:
(i) all nodes get a separate community label; (ii) in a loop over all nodes the community label of the current node is assigned to the nodes that have not changed their label yet and whose cohesion with the current node exceeds the threshold.
One could also start building dendrograms {\cz (tree diagrams frequently used to illustrate the arrangement of the clusters produced by hierarchical clustering)} using simple methods such as  single-linkage clustering or complete linkage clustering \cite{Fortunato10, clustering1, clustering2}. 
While the gap seen in the histogram is a clear indication of a modular network structure, these methods rarely provide optimal community identification.

Moreover, the accuracy of the above outlined method  of complexity $\mathcal{O}(N^2)$ depends on the depth of the gap and the exact threshold value. 

It turns out, however, that by a feedback procedure that modifies the network microstructure while preserving the community superstructure unaltered, node pairs can be completely separated into a low- and a high cohesion group manifesting in an amplified contrast of the cohesion picture [see Fig. \ref{F:boost}(b)]. 

For some networks, including the one in Fig. \ref{F:boost}, these groups can be directly identified as inter- and intra-module pairs. The clear separation of the two groups makes the choice of the threshold value arbitrary (within the limits of the gap) and allows the identification of the communities, i.e., a full reproduction of the truth information.}
\begin{figure*}[htbp]  
\begin{center}
{\includegraphics[width=0.8\linewidth]{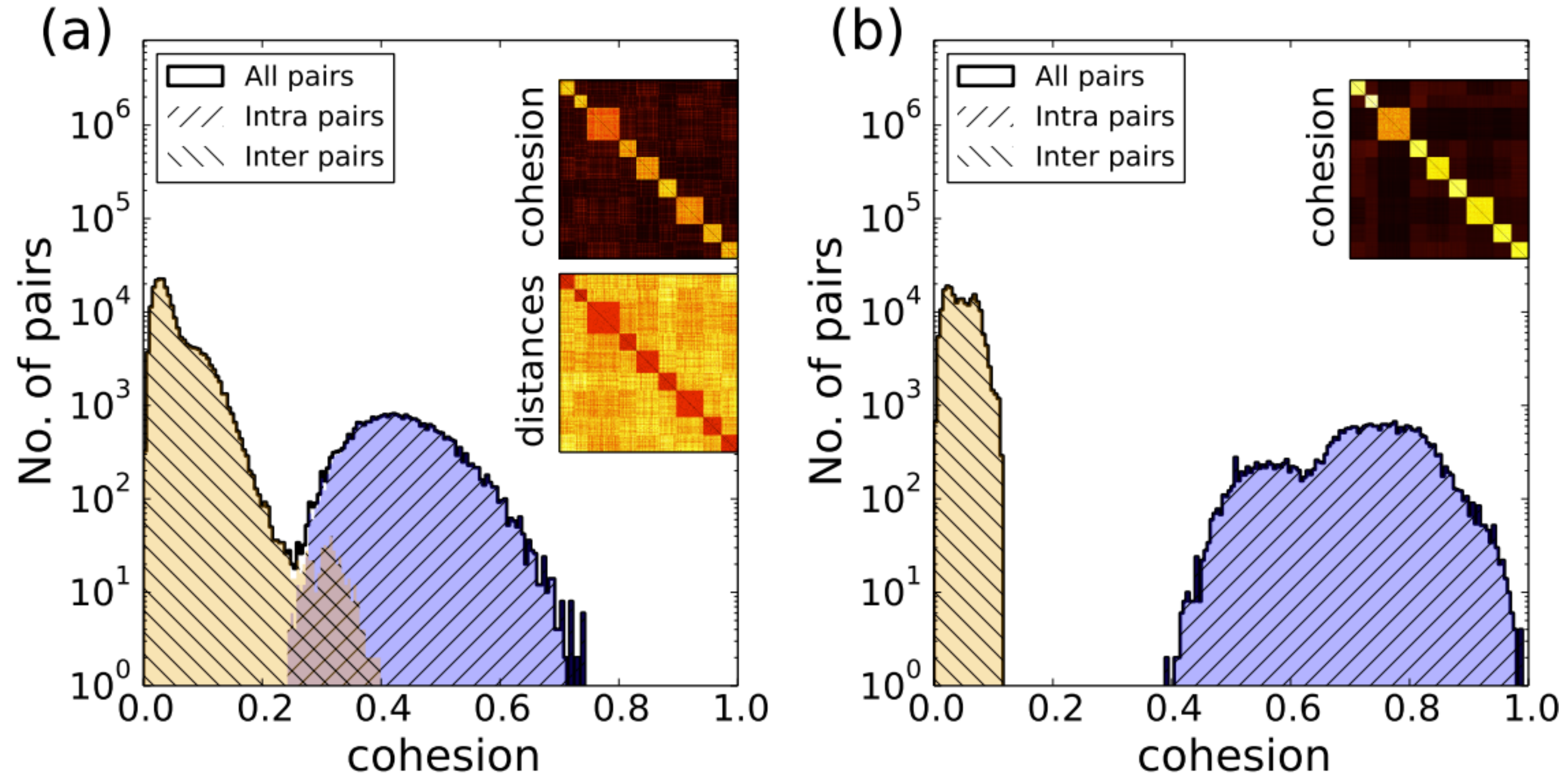}}\,
\caption{ (color online) (a) Cohesion histogram for all node pairs in the cohesion map of an $m=9$ module undirected and unweighted benchmark network: $N=500$, $M=5000$, $g=15$, and $R=3000$ \cite{benchmark1,benchmark2}. 
Intra- and intermodular pairs are identified and colored based on truth information.  
The cohesion and inter-node distance matrices are presented as insets with lighter colors corresponding to higher values. 
(b) Cohesion histogram for the same network using the contrast boosting technique discussed in Sec. \ref{S:boost}}.
\label{F:boost}
\end{center}
\end{figure*}
\begin{figure*}[htbp]  
\begin{center}
{\includegraphics[width=\linewidth]{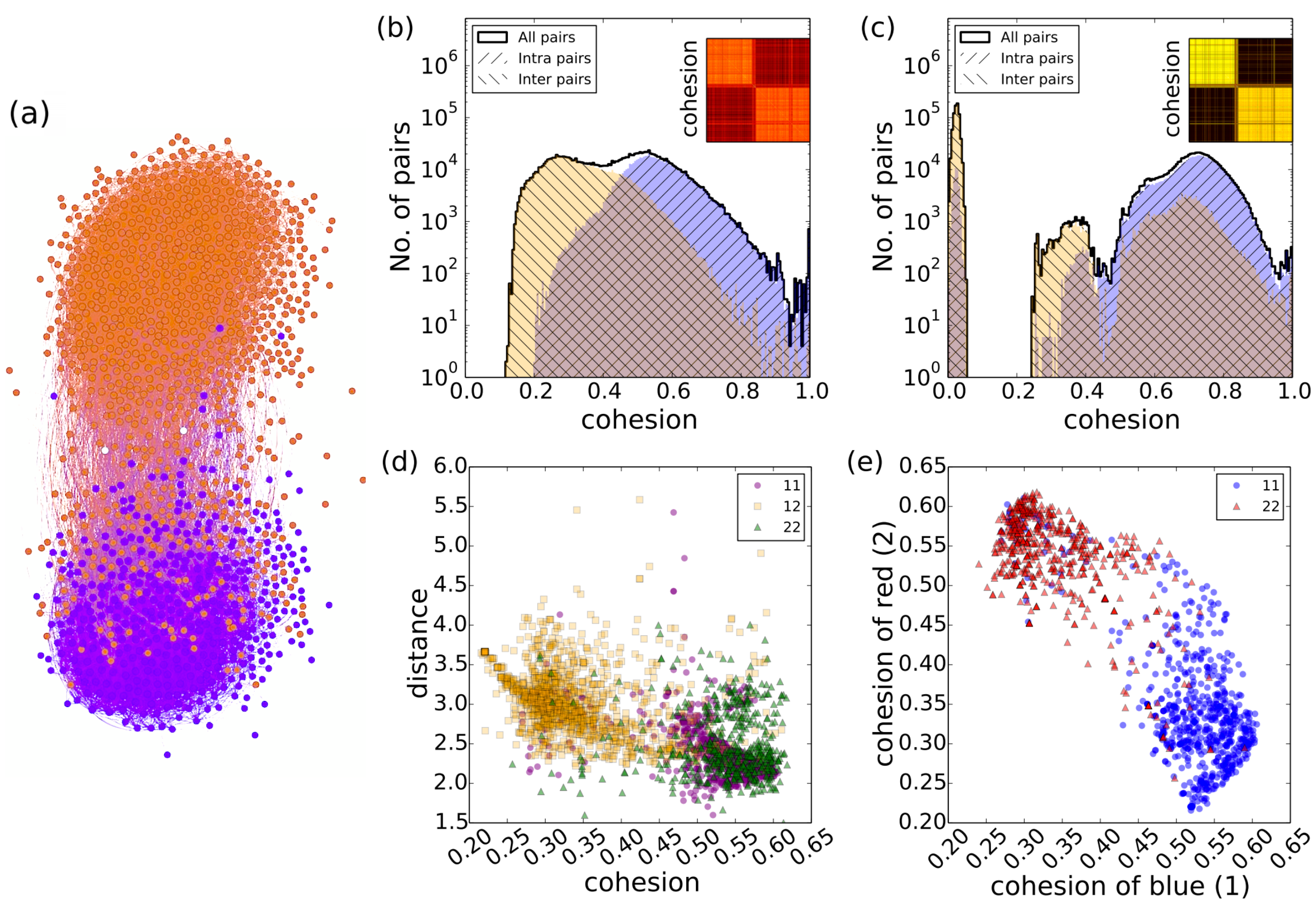}}\,
\caption{ (color online) Network of American political blogs (POLBLOGS) separated into two communities referred to as 1 [violet (darker gray)] and 2 [orange (lighter gray)] by the edge relocation method described in Sec. \ref{S:boost}. (a) The representation of the network in Gephi using the ForceAtlas2 layout algorithm \cite{gephi}. All edges are assumed to have unit length. (b),(c) Voronoi cohesion matrix and histogram  obtained without contrast boost and with the edge relocation methods, respectively (see Sec. \ref{S:boost}). 
{\cz (d)  Average cohesion vs average distance for each node. By average cohesion(distance) of a node we mean the arithmetic mean of the cohesion(distance) matrix elements in the row corresponding to the node but limited to one of the two groups.  Magenta circles (green triangles) represent the average cohesion and distance of a node in group 1(2) to its  own group. Orange squares  correspond to the same quantities {\cz but in relation to the } opposite group. 
 (e) Average cohesion of  a node to group 1 vs the same quantity for group 2. Each node appears as a point colored by its group affiliation (red triangle or blue circle).}}
\label{F:polblogs1}
\end{center}
\end{figure*}
%
%

For achieving the intra-inter segregation of node pairs one can choose from a practically unlimited arsenal of methods. 
Our two successfully implemented methods consist in modifying the network topology based on the cohesion matrix. 
On the nine module benchmark network shown in {\cz Fig. \ref{F:boost}(a)} we applied the following algorithm: 
(1) generate a number of $200$ Voronoi diagrams;
(2) based on the obtained cohesion matrix remove $1\%$ of the worst ``performing" edges; 
(3) repeat the previous two steps $15$ times. 
While in line with the findings of Sec. \ref{S:complexity}, the above numerical values are rather empirical, set on a trial and error basis, mostly determined by the trade-off between low statistical error and computational demand.
The very same method, hereinafter \textit{weak edge removal}, did not produce a complete cohesion segregation for the POLBLOGS network studied in detail in Sec. \ref{S:polblogs}. 
There the procedure was slightly different. 
In step (2) the deleted edges were reinserted between the best performing unlinked node pairs. 
Thus the ratio of intra-module to inter-module connections got boosted while conserving the overall link density. 
This latter scheme, henceforth referred to as \textit{weak edge relocation}, proved to be effective also for benchmark graphs. 
{\cz The next section will demonstrate that } while these contrast boosting methods do not always produce a complete segregation, they can significantly improve community detection.

\section{Real networks}\label{S:real}

{\cz Network clustering} using Voronoi cohesion proved to be successful on benchmark networks.
In order to circumscribe within a more practical context the scope of the stochastic graph-Voronoi tessellation we have considered a number of {\cz widely different} real networks frequently used in the literature for testing novel network measures and methods. {\cz Our conclusions are summarized in Sec. \ref{S:conclusions}.}

\subsection{Political blogs}\label{S:polblogs}
The network of links between American political blogs (POLBLOGS) has $N=1223$ nodes and $M=16 715$ edges \cite{polblogs}. 
Community detection algorithms identify two large groups that we shall refer to by labels 1 and 2 or colors violet (darker grey) and orange (lighter grey), respectively [see Fig. \ref{F:polblogs1}(a)]. {\cz This community information will be used as substitute for truth.}
The plain stochastic graph-Voronoi method as described in Sec.s \ref{S:intro} and \ref{S:overview} produces the cohesion map shown in Fig. \ref{F:polblogs1}(b). 
Though present, the gap in the cohesion histogram is not overly pronounced. 
Applying the contrast boosting technique detailed in Sec. \ref{S:boost} a clear gap is formed allowing the identification of communities [see Fig. \ref{F:polblogs1}(c)]. 
Unlike the benchmark network presented in Sec. \ref{S:overview} here the gap does not ``hermetically'' separate intra- and inter-module node pairs. 

{\cz Since Voronoi tessellation is based on distances  it is only natural to ask} whether the Voronoi cohesion matrix provides more information about community structure than simple graph distances.  In Fig. \ref{F:polblogs1}(d) we plot for each node {\cz its average distance} to the nodes within the same module (magenta circles for module 1, green triangles for module 2)  against the similarly calculated cohesion values. We also plot for each node {\cz its average distance and cohesion to the nodes} in the opposite module (orange squares).  There are 2$N$ points in total in Fig. \ref{F:polblogs1}(d).}
The discrepancy between the separability of intra- and intermodular groups along the two axes demonstrates the superior usefulness of Voronoi cohesion as compared to graph distance. 
By representing each node {\cz in the two dimensional space of their average cohesions  to the two different modules,} the communities become well separated and exhibit specific traits [see Fig. \ref{F:polblogs1}(e)]. Subgroups and individuals can be identified and evaluated in terms of their ``attitude''  in relation to their ``own'' versus ``the others''. 

\subsection{Zachary's karate club}\label{S:karate}
The friendships between the 34 members of the Zachary karate club are captured by a network of 78 links \cite{karate}.

The clusterization of such small networks is tractable even by greedy methods therefore we are only interested whether Voronoi cohesion offers additional insight and how this information is affected by the number of seed nodes.
\begin{figure}[htbp]  
\begin{center}
\resizebox{0.9\columnwidth}{!}
{\includegraphics{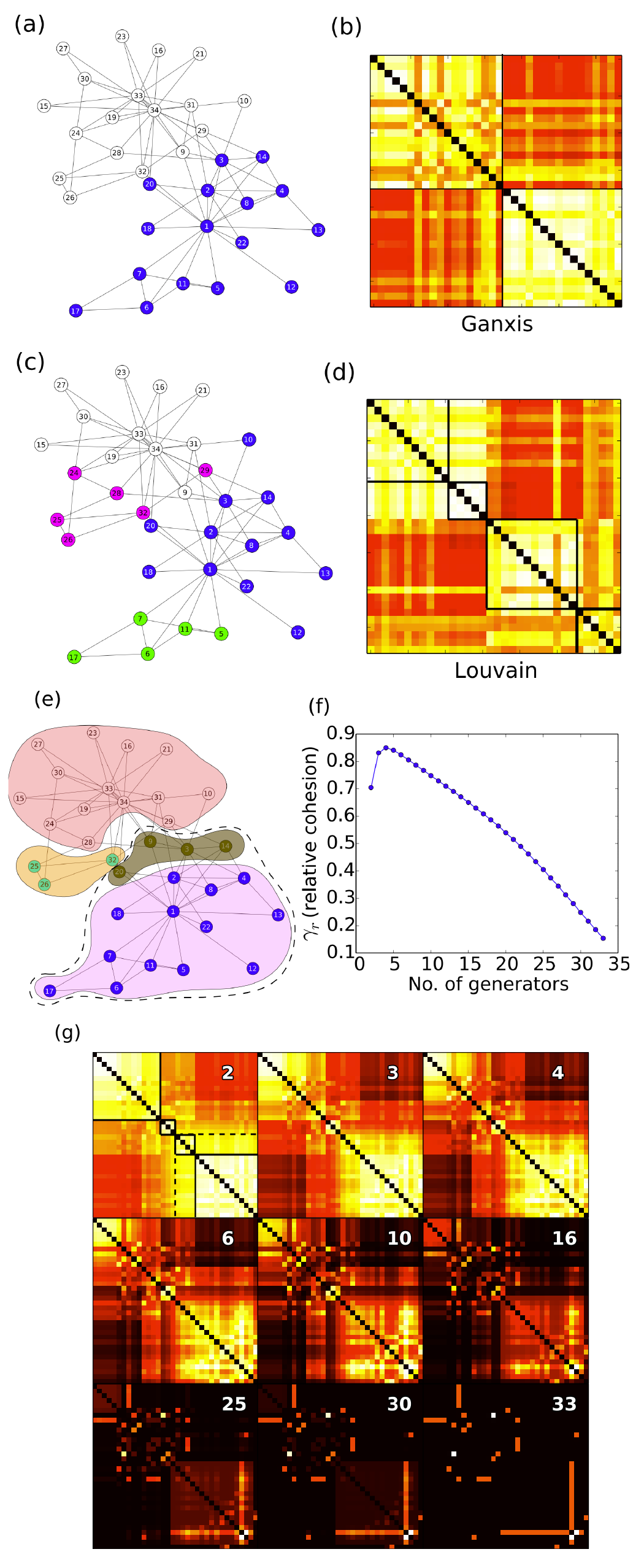}}

\caption{  (color online) Influence of the number of seeds exemplified on a small social network (Zachary's karate club) \cite{karate}. {\cz The layout is generated using the LGL network visualization method \cite{lgl} as implemented in the iGraph software package \cite{igraph}}.
Statistical errors are {\cz eliminated} by considering all possible Voronoi diagrams. 
(a),(c) Topology and modular structure according to the GANXiS \cite{ganxis} and  Louvain \cite{Louvain} algorithms. 
(b),(d) Voronoi cohesion maps for $g=2$ seeds. 
Node ordering is based on the two clusterizations. 
(e) Modular structure established visually based on the Voronoi cohesion map. 
Starting from the map in (b) nodes were rearranged manually so that block diagonality was improved yielding the map in the top-left tile in (g).
(f) Dependence of the relative contrast [see Eq. (\ref{contrast_relative})] on the number of seeds. 
As replacement for truth information the GANXiS clusterization is used [see (a)].
(g) Cohesion matrices for different numbers of seeds. 
Values range from 0 (black) to 1 (white). 
Nodes are ordered as described in (e).
Seed numbers are indicated in the top-right corner of each map.
}
\label{F:karate}
\end{center}
\end{figure}
The statistical error in estimating Voronoi cohesion was {\cz mostly eliminated} by generating all possible tessellations for all seed numbers, $g$, from 2 to 33. 
The GANXiS community detection method \cite{ganxis} identified two clusters [see Fig. \ref{F:karate}(a)]. 
By ordering the nodes based on this clusterization produces a $g=2$ cohesion matrix shown in Fig. \ref{F:karate}(b).
Applying the same procedure with the Louvain community detection \cite{Louvain} four clusters are obtained [see Figs. \ref{F:karate}(c) and \ref{F:karate}(d)].  
{\cz The ordering of the nodes based on the two clusterizations yields  cohesion matrices relatively far from block diagonal. Moreover, the small size of the network enhances the relative importance of individual nodes; therefore, a clustering that is mostly determined by the particularities of the automated method cannot be taken as bona fide.
However, reordering the nodes manually in order to achieve a structure of the cohesion map closer to block diagonal is feasible at this network size and 
can lead to a more expressive community picture [see top-left tile in Fig. \ref{F:karate}(g)] suggesting a more refined community structure outlined in Fig. \ref{F:karate}(e). 
}

The contrast defined in Eq. (\ref{contrast2}) with average intra- and inter-module cohesion values estimated based on the GANXiS clusterization reaches a maximum for generator size of $g\approx 25$. 
However, the compromised cohesion ``picture" [see Fig. \ref{F:karate}(g)] again demonstrates the unsuitability of this measure for quantifying clustering. 
On the other hand the relative contrast from Eq. (\ref{contrast_relative}) yields an optimal seed number of $g\approx 4$ [see Fig. \ref{F:karate}(f)] value more in line with the naked eye approach. 

We observe from Fig. \ref{F:karate}(g) that while high generator number is not appropriate for circumscribing modules it reveals some lower scale particularities of groups and individual nodes.

\subsection{Caenorhabditis elegans}\label{S:elegans}
{\cz The C. elegans is a primitive multicellular organism (roundworm) of great interest in life sciences and adjacent fields.}
Its neural network consists of 297 nodes and 2359 edges \cite{celegans}. 
The Louvain clustering \cite{Louvain} algorithm finds five modules and attains a modularity measure of 0.387 (see the Appendix). 

\begin{figure*}[htbp]  
\begin{center}
{\includegraphics[width=0.8\linewidth]{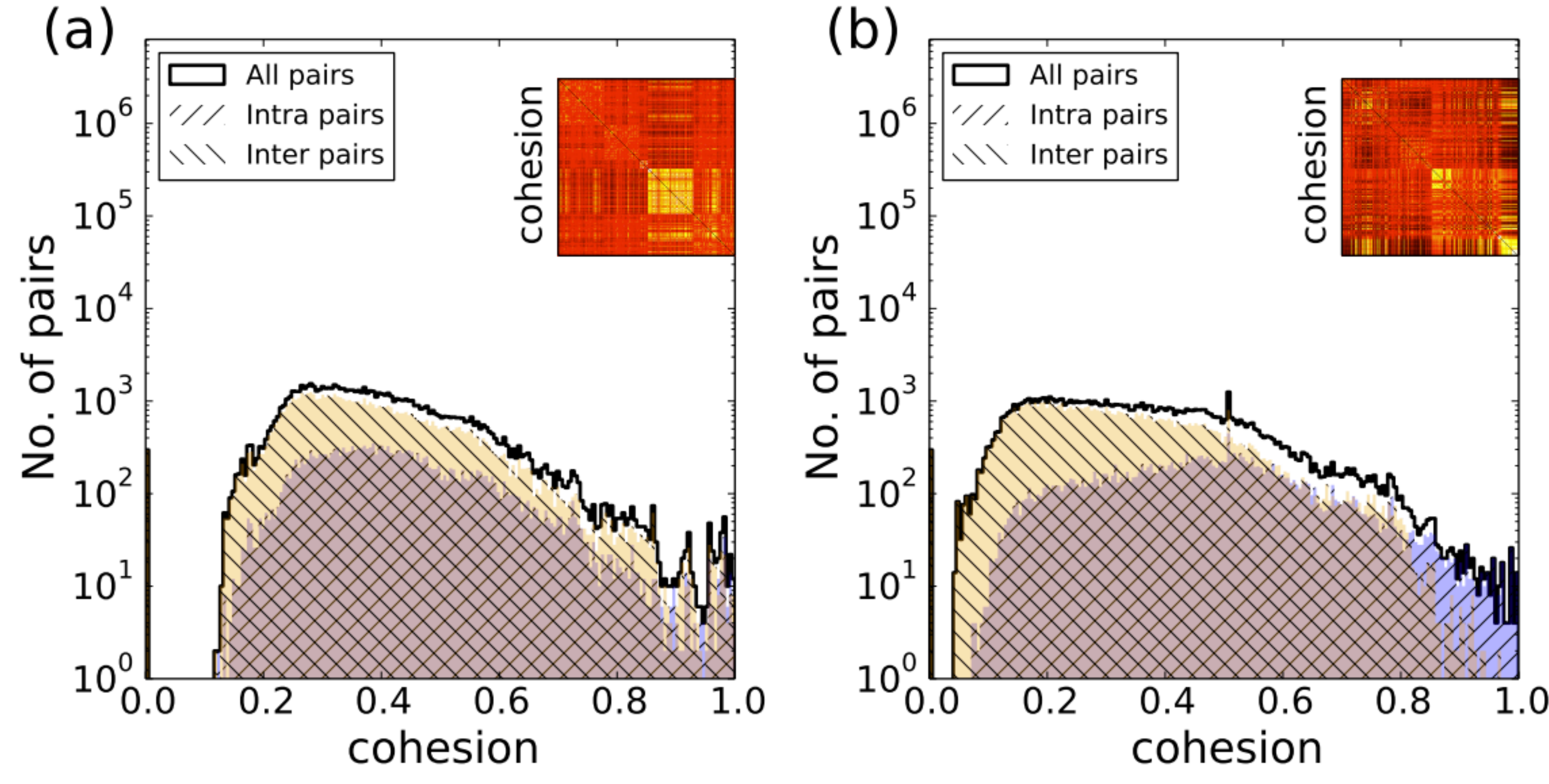}}\,
\caption{ (color online) Neural network of the Caenorhabditis elegans. Voronoi cohesion matrix and histogram yielded by the (a) plain (no contrast boosting)  and the (b) edge relocation methods, respectively.  We can observe a correction in the contrast of the cohesion matrix in (b) compared to (a) (see insets), but the histograms show that very little difference was achieved in terms of separating the inter- and intrapairs. {\cz Surrogate truth information (see the Appendix) was provided by the Louvain clusterization method \cite{Louvain}. Node ordering in the two insets is also based on this information.} }
\label{F:celegans1}
\end{center}
\end{figure*}

The contrast boosting methods described in Sec. \ref{S:boost} yielded a modularity measure of only 0.29. 
In Figs. \ref{F:celegans1}(a) and \ref{F:celegans1}(b) we can see the results of a simple and edge relocated graph-Voronoi tessellation. 
None of the two managed to produce the expected separation gap in the cohesion histogram. 
However, the  {\cz edge relocation} mechanism described in Sec. \ref{S:boost} resulted in a modified network that once clustered by Louvain gave a modularity of 0.404 when the community information was applied on the original network. 
 The result was achieved with 15 seeds and the cohesion was boosted the same way (described previously) in every 400 iterations.
This finding suggests that whenever the stochastic graph Voronoi method cannot directly offer good quality clustering it may still be used in tandem with robust third party methods for optimal results. 

\section{Conclusions}\label{S:conclusions}
Methods that shed light on previously unexplored cross sections of networks are likely convertible into valuable applications.  Prompted by an almost inadvertent observation we explored some of the theoretical and practical aspects of the interaction between the modular structure of networks and the properties of their random graph-Voronoi tessellation. 

The node-pair  measure we called Voronoi cohesion, reveals inter-node relationships that otherwise cannot be captured by the adjacency or the inter-node distance matrices. 
In contrast with other node-similarity measures \cite{similarity,Fortunato10,newman04} Voronoi cohesion provides information based on the global structure of the network.

We have shown that this type of information overlaps with community rapports. 
As such it can be employed for clustering networks with well contoured community structure as demonstrated on benchmark networks and a few real networks.

Due to the recent interest in the large scale structures of networks our initial effort went into exploring the mathematical background of the Voronoi cohesion vs. modularity relationship. 
The contrast of the cohesion matrix defined in Eq. (\ref{contrast2}) and the relative contrast introduced in Eq. (\ref{contrast_relative}) were used as quality measures of this relationship. 
The relatively simple analytical model of two, large ER-type modules partitioned into two graph-Voronoi ``cells" grasps the essentials of the influence of link densities on the cohesion picture (see Sec. \ref{S:analytical}). 
Beyond the basics several aspects surface such as convergence rate or accuracy and the phenomenon exhibits an intricate interplay between a number of parameters including but not limited to network size, number of modules, number of generator nodes, and modularity. 
The role of the only important adjustable parameter, the number of generator nodes (cell centers), is clarified based on the model of loosely connected multimodular networks (see Sec. \ref{S:analyticalextreme}). 
The main conclusion, namely for an optimal contrast the number of generators should be slightly in excess of the number of modules, was proven valid for all studied networks, i.e., well beyond the limits of the model. 
The proposed analytical approach could go a long way in exploring some of these interactions, nonetheless, we preferred simulations to stretching the limits of the method. 
{\cz These studies presented  in Sec. \ref{S:complexity} revealed that the necessary statistical sample size (number of repeats) is relatively low, in the hundreds, and is barely influenced by the size of the network.}

The perfect clusterization of some benchmark networks necessarily asked for testing the limits of the method's community detection capabilities. 
While the Voronoi cohesion matrix and the extracted relative contrast can be used to visually and quantitatively compare different clusterizations additional procedures are required to make it suitable as a stand alone community detector. In Sec. \ref{S:boost} we build on a basic procedure that converts the information residing in the cohesion matrix into community labels for each node. The method's low complexity automatically curbs the range of networks it can be applied on. Acceptable performance is obtained only for strongly modular cases. This constraint can be circumvented by different cohesion-to-topology feedback techniques that boost the modularity of network (see Sec. \ref{S:boost}). For the POLBLOGS network (see Sec. \ref{S:polblogs}) the forced separation of node pairs into strongly and weakly interacting groups proved to be a working strategy yielding a clusterization with a modularity measure identical to that obtained by Louvain.  
 For real networks as small as Zachary's karate club (see Sec. \ref{S:karate}) or those with less standard modular structure, e.g., the C. elegans neural network (see Sec. \ref{S:elegans}) the tested techniques failed to produce modularity measures similar to or higher than Louvain.  On the other hand, the  {\cz edge relocation} technique described in Sec. \ref{S:boost} can be successfully applied ``on top of" Louvain allowing further improvements in terms of Newman's modularity measure \cite{newman04}.

However, in our view the primary significance and utility of Voronoi cohesion consists in offering a continuous local measure extracted in view of the global context. 
This individual node-node relationship measure can be extended to cover node-group and group-group affinities. It can be applied irrespective of the type and level of network modularity. Some of these features were illustrated through the example of the POLBLOGS social network discussed in Sec. \ref{S:polblogs} [see Fig. \ref{F:polblogs1}(e).] {\cz As mentioned in the Introduction the fixed generator model studied in \cite{Deritei14} may have relevance in understanding the implications of the right choice of candidate representatives in the social network of a heterogeneous electorate. In a similar social context our randomized generator sets correspond to candidates that are arbitrary individuals and Voronoi cohesion reflects the likelihood for any two voters to express similar options. }

We also consider that Voronoi cohesion has the potential to play an important role in future community detection algorithms {\czz allowing for a number of extensions and generalizations. For example, simple geodesic distance could be calculated based on some local measure such as the edge clustering coefficient \cite{Deritei14} or replaced by random walk distance \cite{Schaub16}}.

In the meantime, combined with  modularity boosting techniques it can be used for assisting and supervising third party community detection methods. As suggested by Fig. \ref{F:multilevel} networks with hierarchical and overlapping community structures, though unexplored in the present paper,  might be another promising avenue toward important applications of stochastic graph-Voronoi tessellation.

\section*{Acknowledgments}
This work was supported by the Romanian National Authority for Scientific Research (CNDI-UEFISCDI) Grants No. PN-II-PT-PCCA-2011-3.2-0895 and No. PN-III-P2-2.1-BG-2016-0252. I.P. and Z.I.L acknowledge the support of Bethlen G\'abor Alap through Collegium Talentum, Hungary, contract No. TB-03/258/0/2015. M. E.-R.  received funding from the European Union’s Horizon 2020 research and innovation programme under Grant Agreement No 668863 (SyBil-AA). The authors are grateful to Zolt\'an N\'eda for the enlightening discussions and useful suggestions. 

\section*{APPENDIX}
\noindent
{\cz
\textbf{Vertex}: node of a graph\\
\textbf{Link, edge}: connection between two nodes of a graph\\
\textbf{Erd\H{o}s-R\'enyi (ER) graph}: random graph characterized by a single link density parameter that specifies the probability of any pair of vertices to be connected \cite{randomgraphs}.\\
\textbf{Community, cluster, module}: subgraph with a pronounced internal connectivity \cite{Fortunato10} \\
\textbf{Benchmark network}: algorithm generated network with a priori set parameters. They are often used for testing graph related methods. When testing community detection methods benchmark networks are built by  connecting some a priori defined communities. \cite{benchmark1, benchmark2}\\
\textbf{Truth, ground truth}: the information available prior to and independent of the result of the community detection procedure. For instance, the community information, i.e, node-to-community mapping, for \textbf{benchmark graphs}. This is regularly compared to the output of the community detection for assessing its performance.\\
\textbf{Modularity measure}: designed to quantify the strength of division of a network into \textbf{modules} \cite{newman04}. It takes the network and a node-to-community map as input, returns a value in the [-1/2, 1) range. Used also for assessing community detection methods when no \textbf{truth} information is available. \\
\textbf{Louvain, GANXiS}: mainstream community detection methods \cite{Louvain, ganxis}\\
\textbf{Graph Voronoi diagram}: given a set of $g\in\mathbb{N}^*$ vertices (\textbf{generator} nodes) in a graph the latter is separated into $g$ disjoint subgraphs or \textbf{cells} such that all nodes in a \textbf{cell} are closest in terms of some graph metric to the associated \textbf{generator} node \cite{graphvoronoi}\\
\textbf{Cell, Voronoi cell}: subgraph containing those nodes that are closest in terms of some graph metric to the \textbf{generator} node associated with the cell. 
\\
\textbf{Generator, generator node, center, seed}: node associated with a \textbf{Voronoi cell}.  The other vertices in the cell are closest, from the set of all generators, to this particular node. 
\\
\textbf{Cohesion, Voronoi cohesion}: the probability for two nodes to belong to the same \textbf{Voronoi cell} in the statistical ensemble of \textbf{graph Voronoi diagrams} with randomly chosen \textbf{generator} nodes\\
\textbf{Repeats}: the size of the statistical ensemble (number of samples)\\
\textbf{Bridge density}: the ratio of the number of existing links to the number of possible links between two \textbf{modules}\\

 }

\end{document}